\documentclass[aps,prl,reprint,superscriptaddress,nofootinbib,floatfix]{revtex4-2}
\usepackage{amsmath,amssymb,bm}
\usepackage{graphicx}
\usepackage[colorlinks=true,citecolor=blue,linkcolor=blue,urlcolor=blue]{hyperref}
\newcommand{\maybegraphics}[2][]{%
  \IfFileExists{#2}{\includegraphics[#1]{#2}}{%
    \fbox{\begin{minipage}{0.92\columnwidth}\centering Missing figure file: \texttt{\detokenize{#2}}\end{minipage}}%
  }%
}

\begin{document}

\title{Kinetic Criticality in Linker-Mediated Colloidal Aggregation}

\author{Alexei V. Tkachenko}
\email{oleksiyt@bnl.gov}
\affiliation{Center for Functional Nanomaterials, Brookhaven National Laboratory, Upton, New York 11973, USA}
\author{Soojung Lee}
\affiliation{Department of Chemical Engineering, Columbia University, New York, New York 10027, USA}
\author{Zohar A. Arnon}
\affiliation{Department of Chemical Engineering, Columbia University, New York, New York 10027, USA}
\affiliation{Avram and Stella Goldstein-Goren Department of Biotechnology, Ben-Gurion University of the Negev, Beer Sheva 8410501, Israel}
\author{Oleg Gang}
\affiliation{Center for Functional Nanomaterials, Brookhaven National Laboratory, Upton, New York 11973, USA}
\affiliation{Department of Chemical Engineering, Columbia University, New York, New York 10027, USA}
\affiliation{Department of Applied Physics and Applied Mathematics, Columbia University, New York, New York 10027, USA}
\affiliation{Center for Nanomedicine, Institute for Basic Science and Department of Nano Biomedical Engineering, Yonsei University, Seoul 03722, Republic of Korea}

\date{\today}

\begin{abstract}
Linker-mediated aggregation plays an important role in modern nanoscience. We demonstrate that it departs sharply from classical Smoluchowski kinetics because cluster reactivity evolves during growth. Combining theory with DNA-linked gold-nanoparticle experiments, we establish kinetic critical point controlled by linker abundance. Below threshold, active linkers are depleted and growth arrests; above threshold, clusters accumulate reactive sites, self-accelerate, and cross over to diffusion-limited coarsening. Experiments verify the predicted arrest, accelerated growth, and scaling collapse.
\end{abstract}

\maketitle

Colloidal aggregation is one of the foundational  problems in nonequilibrium statistical mechanics and soft condensed matter pohysics. In the classical Smoluchowski picture of diffusion-limited coagulation, cluster encounters are controlled by transport, and the mean cluster mass grows approximately linearly in time, $M(t)\sim t$ \cite{Smoluchowski1916,Leyvraz2003Scaling}. The experimentally observed hydrodynamic radius grows even slower, 
\[
R_h(t)\sim M^\nu \sim t^\nu ,
\]
where $\nu=1/d_f$, and $d_f$ is fractal dimensionality of the aggregate. In three-dimensional diffusion-limited cluster aggregation, simulations and colloid experiments give $d_f\simeq 1.7$--$1.9$, corresponding to $\nu\simeq 0.53$--$0.59$ \cite{Meakin1984DLCA3D,Weitz1984GoldFractals,Lin1989Universality}. Thus, in conventional aggregation, the dynamics slows as aggregates coarsen.

Here we discuss linker-mediated aggregation and demonstrate how it leads to a qualitative revision of  the classical picture. Linker-mediated binding is a common motif in programmable soft matter and biomolecular recognition. In DNA-mediated colloidal assembly, mobile or soluble DNA linkers bind to particle surfaces and then bridge neighboring particles, producing interactions controlled by linker concentration, sequence specificity, and valency \cite{Xiong2009DNALinkerPhase,AngiolettiUberti2012Reentrant,Varilly2012GeneralTheory,DiMichele2013Multistep,AngiolettiUberti2014MobileLinkers,Zhang2015SelectiveTransformations,AngiolettiUberti2016Guide,AngiolettiUberti2019Understanding,Cui2022MicroscopicInteractions,Lowensohn2019PhaseBehavior,Rogers2020MeanFieldLinker,Xia2020MobileDNA}. The same physical motif also appears in protein- and ligand-mediated recognition, where multivalent ligand--receptor interactions underlie selective binding, immunoassays, and aggregation-based optical biosensors \cite{Mammen1998Polyvalent,MartinezVeracoechea2011Multivalency,Thanh2002AggregationImmunoassay,Tsai2005ProteinInteractions,Liu2009ScFvGoldImmunoassay,Iarossi2018ColorimetricImmunosensor}.

In the case of linker-mediated aggregation, cluster reactivity is not fixed once and for all by particle geometry. Instead, it evolves during aggregation: as clusters merge, active linkers are redistributed among fewer aggregates, so larger clusters can carry more reactive sites. This creates a positive feedback between cluster growth and cluster reactivity, producing a self-accelerating linker-limited regime. This mechanism is distinct from both diffusion-limited cluster aggregation (DLCA) and standard reaction-limited cluster aggregation (RLCA), which provide the classical limiting descriptions of irreversible colloidal aggregation \cite{Meakin1984DLCA3D,Weitz1984GoldFractals,Lin1989Universality,Leyvraz2003Scaling,Lin1990RLCA}. In DLCA, clusters stick upon contact and form ramified fractal aggregates; in ordinary RLCA, a reduced sticking probability or activation barrier slows aggregation, but the reactive valence of a cluster does not increase with its mass. Here, by contrast, the effective valence evolves dynamically: clusters can become increasingly reactive as they grow, aggregation accelerates rather than slows. Alternatively, if the system exhausts its available active linkers, growth arrests at finite cluster size. The two regimes are separated by a kinetic critical point.

This kinetic criticality connects naturally to gelation, patchy particles, and linker-mediated assembly. In Flory--Stockmayer theory, network formation is controlled by bonds between units of prescribed functionality, and gelation occurs when the branching connectivity exceeds a critical value \cite{Flory1941Gelation,Stockmayer1944GeneralCrosslinking}. In patchy-particle and limited-valence models, this idea reappears in coarse-grained form: percolation, gelation, and equilibrium phase behavior are governed by valence, bond probability, and reversibility \cite{Sciortino2010PrimitivePatchy,Bianchi2011PatchyReview,SciortinoZaccarelli2017EquilibriumGels}. The linker-mediated system studied here is different because the effective functionality is not fixed in advance. Linkers act as bifunctional bridges, while particles and clusters are multivalent objects whose instantaneous reactivity depends on adsorbed linker number, linker consumption, cluster growth, and contact-induced surface exclusion \cite{MartinezVeracoechea2011Multivalency,Howard2021WertheimDoubleBond}. Thus the classical gelation/percolation distinction between finite clusters and an infinite network appears here as a kinetic fate: below threshold, active linkers are depleted and growth arrests at finite size, whereas above threshold, enough active linkers remain to sustain unbounded aggregation.

Previous treatments of linker-mediated aggregation have shown that linker-to-particle ratio, linker diffusivity, and cluster composition strongly influence aggregation outcomes \cite{Antunes2019OptimalLinkers,Tavares2020LinkerSmoluchowski}. In particular, Tavares et al. formulated generalized Smoluchowski equations in which clusters are classified by both particle number and linker content \cite{Tavares2020LinkerSmoluchowski}. Our approach differs in the aggregation kernel: rather than assuming a finite-valence or phenomenological association rate, we derive the rate from patch-limited capture, treating the local reaction problem in the spirit of Berg--Purcell diffusion to small targets and related partially absorbing surface problems \cite{BergPurcell1977,Shoup1982LigandBinding,Zwanzig1990DiffusionToTraps}. This construction captures both early linker-limited amplification, where cluster reactivity increases with size, and the later crossover to diffusion-limited coarsening when transport becomes limiting.

\begin{figure}[t]
\centering
\maybegraphics[width=\columnwidth]{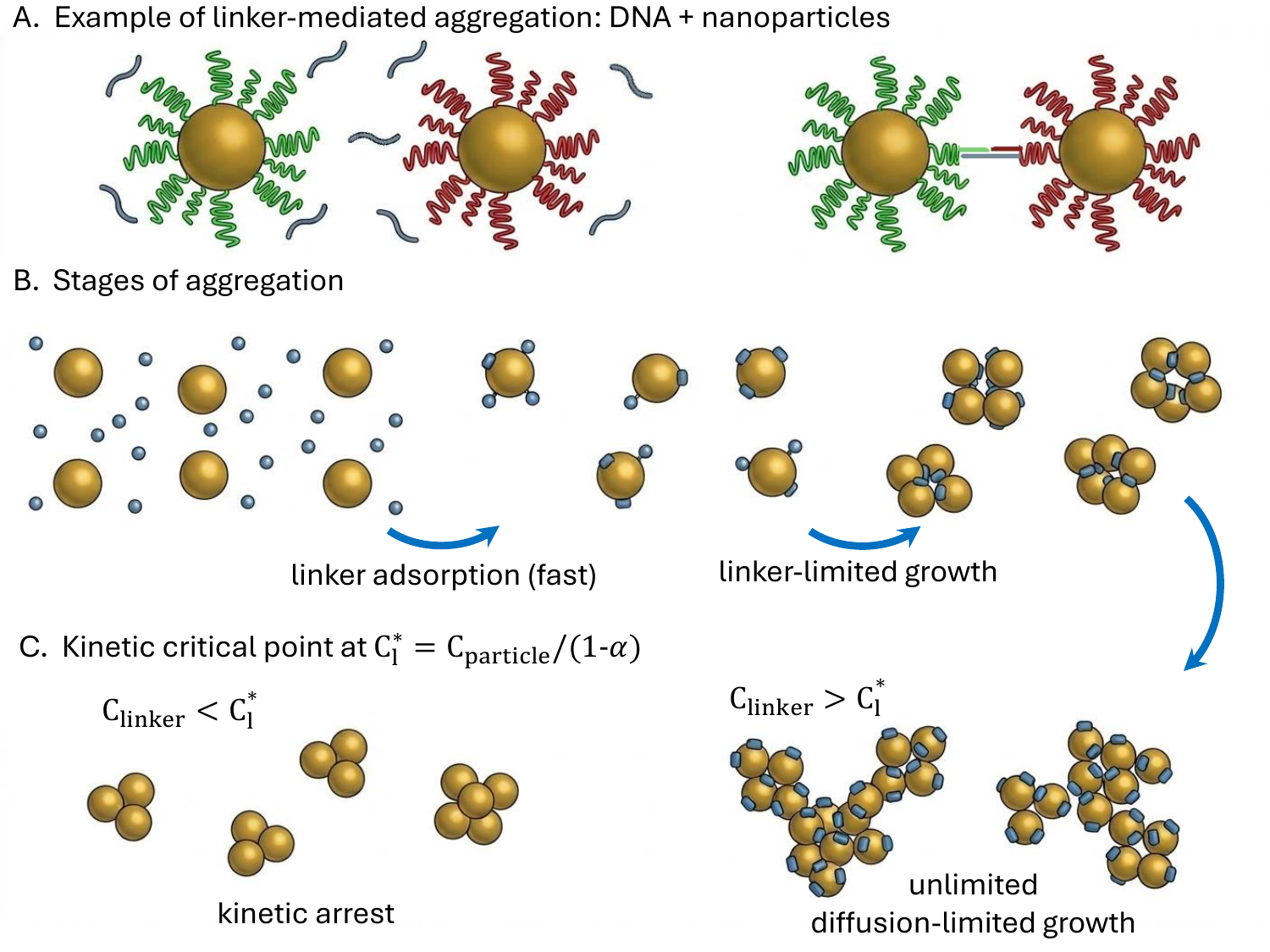}
\caption{Schematic of kinetic criticality in linker-mediated aggregation. (a) DNA-functionalized nanoparticles interact through soluble single-stranded DNA linkers. (b) Free linkers first adsorb rapidly to particle surfaces, creating active patches; subsequent cluster--cluster encounters consume active linkers and block an effective surface fraction $\alpha$ per interparticle contact. (c) With particle concentration $C_p$ and linker concentration $C_l$, the conservation law in Eq.~(\ref{eq:balance}) gives the critical concentration $C_l^*=C_p/(1-\alpha)$. For $C_l<C_l^*$, $c_l\to0$ and growth arrests at finite mass. For $C_l>C_l^*$, active linkers remain available and aggregation crosses from linker-limited acceleration to diffusion-limited coarsening. In the symmetric binary experiments, $C_p$ denotes the concentration of either complementary particle population.}
\label{fig:schematic}
\end{figure}

We denote by $c(t)$ the cluster concentration, by $M=C_p/c$ the mean cluster mass, and by $c_l(t)$ the concentration of active adsorbed linkers capable of forming intercluster bridges.  While the linkers a typically free initially, we assume their initial adsorption at the particles to be nearly irreversible and much faster than the aggregation itself.  

The standard  diffusion-limited coagulation is described by 
\begin{equation}
\dot c=-\kappa_0 c^2,
\qquad
M(t)=1+\kappa_0 C_p t,
\qquad
R_h\simeq r_p M^\nu .
\label{eq:classical}
\end{equation}
For comparable clusters, the Stokes--Einstein relation gives $\kappa_0=8\times10^3RT/(3\eta)$ in ${\rm M}^{-1}{\rm s}^{-1}$, or about $7~{\rm nM}^{-1}{\rm s}^{-1}$ in water at room temperature.

In the linker-mediated problem, each merger between clusters   and reduces the total surface area available for binding  by amount  $4\pi r_p^2 \alpha$. An upper bound estimate,  $\alpha\gtrsim 1/2$, can be obtained  by considering formation of a single dimer that results in blockage of $1/2$ of a single particle surface area.  More generally, $\alpha$ is  an effective excluded-surface parameter that can be reduced by overlap between blocked regions of  linker flexibility. Each bond reduces $c$ by one cluster, consumes one  linker, and additionally reduces the number of active linkers by $\alpha C_l/C_p$ due to area blockage.  This bookkeping gives  
\begin{equation}
 c_l(t)-\gamma c(t)=(1-\alpha)C_l-C_p, 
\label{eq:balance}
\end{equation}
Here $\gamma=1+\alpha\frac{C_l}{C_p}$. The sign of the right-hand side fixes the long-time fate, depending wether linker concentration $C_l$ is above or below critical value, 
\begin{equation}
C_l^*=\frac{C_p}{1-\alpha} .
\label{eq:clstar}
\end{equation}
The critical point  separates two qualitatively distinct regimes, as illustrated schematically in Fig.~\ref{fig:schematic}C. For $C_l<C_l^*$, so the active-linker population is eventually exhausted and aggregation arrests at the finite cluster mass.  For $C_l>C_l^*$, a finite population of active linkers can be sustained and growth can continue indefinitely.

This threshold has a simple interpretation in the language of Flory--Stockmayer branching theory \cite{Flory1941Gelation,Stockmayer1944GeneralCrosslinking}. For particles of fixed functionality $f$, the Flory--Stockmayer percolation occurs when number of linkers per particle satisfies
\begin{equation}
N_l\left(1-\frac{1}{f}\right)>1.
\label{eq:FScriterion}
\end{equation}
In the present model, $N_l=C_l/C_p$. 
Comparison with \eqref{eq:FScriterion} shows that the kinetic critical point corresponds to a Flory--Stockmayer-like branching threshold with effective colloidal functionality
\begin{equation}
f_{\rm eff}=\frac{1}{\alpha}.
\end{equation}

To further explore the kinetic criticality, define
\begin{equation}
\Lambda=\frac{C_l/C_l^*-1}{\gamma},
\qquad
s=|\Lambda|M .
\label{eq:lambda_s}
\end{equation}
Equation~(\ref{eq:balance}) becomes
\begin{equation}
 c_l=\gamma C_p\left(M^{-1}+\Lambda\right).
\label{eq:clM}
\end{equation}
Thus $\Lambda<0$ gives an arrested mass $M_{\rm max}=|\Lambda|^{-1}$, whereas $\Lambda>0$ permits indefinite growth. The characteristic mass and hydrodynamic size diverge as
\begin{equation}
M^*=|\Lambda|^{-1},
\qquad
R_h^*=r_p |\Lambda|^{-\nu} .
\label{eq:scales1}
\end{equation}

\begin{figure}[t]
\centering
\maybegraphics[width=\columnwidth]{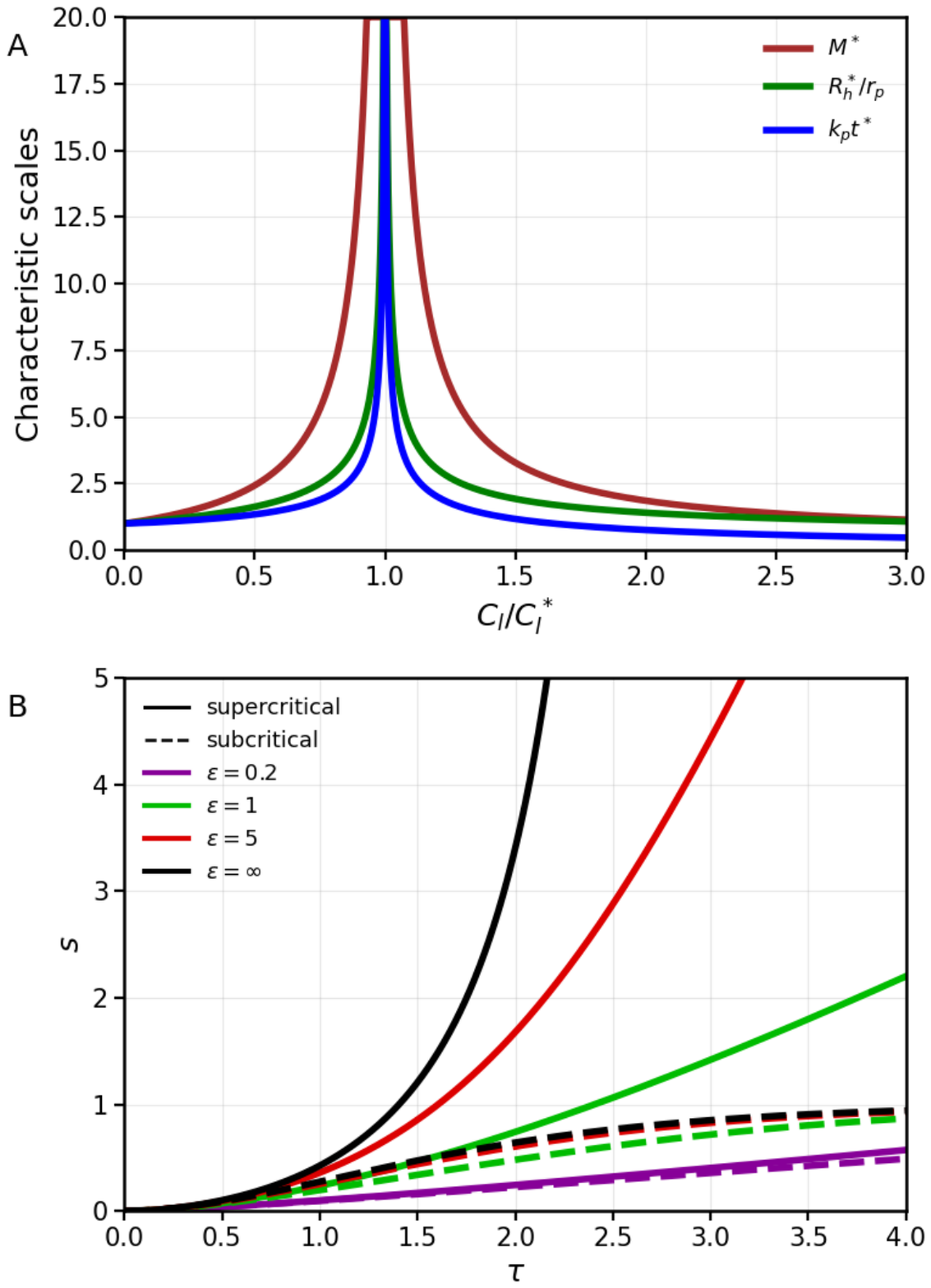}
\caption{Kinetic criticality in linker-mediated colloidal aggregation.
(a) DNA-functionalized nanoparticles bind through short single-stranded DNA linkers, forming interparticle bridges.
(b) Free linkers first adsorb rapidly to particle surfaces, creating active binding sites, followed by linker-mediated cluster growth.
(c) Below $C_l^*=C_p/(1-\alpha)$, active linkers are exhausted and growth arrests at finite cluster size; above threshold, aggregation continues and crosses over to diffusion-limited coarsening.}
\label{fig:master}
\end{figure}

% The rate law follows by adding a diffusion-limited resistance and a patch-limited surface-reaction resistance. Treating each active linker as a reactive patch of effective length $\xi$ gives
% \begin{equation}
% \kappa_r\simeq \kappa_0\frac{\xi}{R_h}\frac{c_l C_p}{c^2},
% \label{eq:kr}
% \end{equation}
% and hence
% \begin{equation}
% \dot c=-\frac{c^2}{\kappa_0^{-1}+\kappa_r^{-1}}
%        =-\frac{\kappa_0\xi C_p c_l}{R_h+\xi C_p c_l/c^2} .
% \label{eq:rate}
% \end{equation}
% The microscopic length $\xi$ contains the linker size, activation free energy, surface blocking, and the compositional factor $z$; for the binary complementary system used experimentally, $z=2$.

We model each adsorbed linker as a circular reactive patch of effective radius $\xi_l\simeq l\exp(-w/k_BT)$, where $w$ is an effective activation free energy. The local patch-to-particle capture problem can be solved by mapping anisotropic diffusion near the patch to an equivalent electrostatic problem for an absorbing disk, as shown in the  Supplementary Materials (SM).  After averaging over the cluster population and absorbing order-unity geometric factors into an effective reaction length $\xi\sim \xi_l$, the reaction-limited aggregation constant becomes
\begin{equation}
\kappa_r \approx \frac{\xi\kappa_0}{R_h}\frac{c_l(t)C_p}{c^2}.
\label{eq:kr}
\end{equation}
The full aggregation kinetics follows from the interplay between diffusion and surface reactivity,
\begin{equation}
\dot{c}
=
-\frac{c^2}{\kappa_0^{-1}+\kappa_r^{-1}}
=
-\frac{\kappa_0\xi C_p c_l}{R_h+\xi C_p c_l/c^2}.
\label{eq:rate}
\end{equation}
This expression is closely related to the Berg--Purcell description of diffusion to small reactive targets \cite{BergPurcell1977}, with the key difference that here both the number of targets (active linkers) and the effective target size evolve dynamically during coarsening.

Introducing
\begin{equation}
 t^*=\frac{r_p}{\xi\gamma\kappa_0 C_p |\Lambda|^\nu}
      \equiv\frac{1}{k_p\gamma |\Lambda|^\nu},
\qquad
\tau=\frac{t}{t^*},
\label{eq:tstar}
\end{equation}
with $k_p=\xi\kappa_0 C_p/r_p$, Eq.~(\ref{eq:rate}) reduces to
\begin{equation}
\frac{ds}{d\tau}=\left[\frac{s^{\nu-1}}{1\pm s}+\frac{1}{\epsilon}\right]^{-1},
\qquad
\epsilon=\frac{r_p}{\xi\gamma}|\Lambda|^{1-\nu} .
\label{eq:scaled}
\end{equation}
The upper sign corresponds to $C_l>C_l^*$ and the lower sign to $C_l<C_l^*$. The critical point is therefore accompanied by $t^*\sim |\Lambda|^{-\nu}$, in addition to the diverging mass and size scales in Eq.~(\ref{eq:scales1}).

Equation~(\ref{eq:scaled}) has closed implicit solutions. In the supercritical regime,
\begin{equation}
\tau_+(s)=\frac{s}{\epsilon}+B_{s/(1+s)}(\nu,1-\nu),
\label{eq:tauplus}
\end{equation}
where $B_z(a,b)$ is the incomplete beta function. In the subcritical regime,
\begin{equation}
\tau_-(s)=\frac{s}{\epsilon}+B_s(\nu,0),
\qquad 0\le s<1 .
\label{eq:tauminus}
\end{equation}
The logarithmic divergence of Eq.~(\ref{eq:tauminus}) as $s\to1^-$ is the kinetic arrest at $M=M^*$. In physical variables, both the limiting mass and the relaxation time grow near threshold, so slightly subcritical and slightly supercritical samples can both appear nearly stationary over a finite experimental window.

\begin{figure}[h!]
\centering
\maybegraphics[width=\columnwidth]{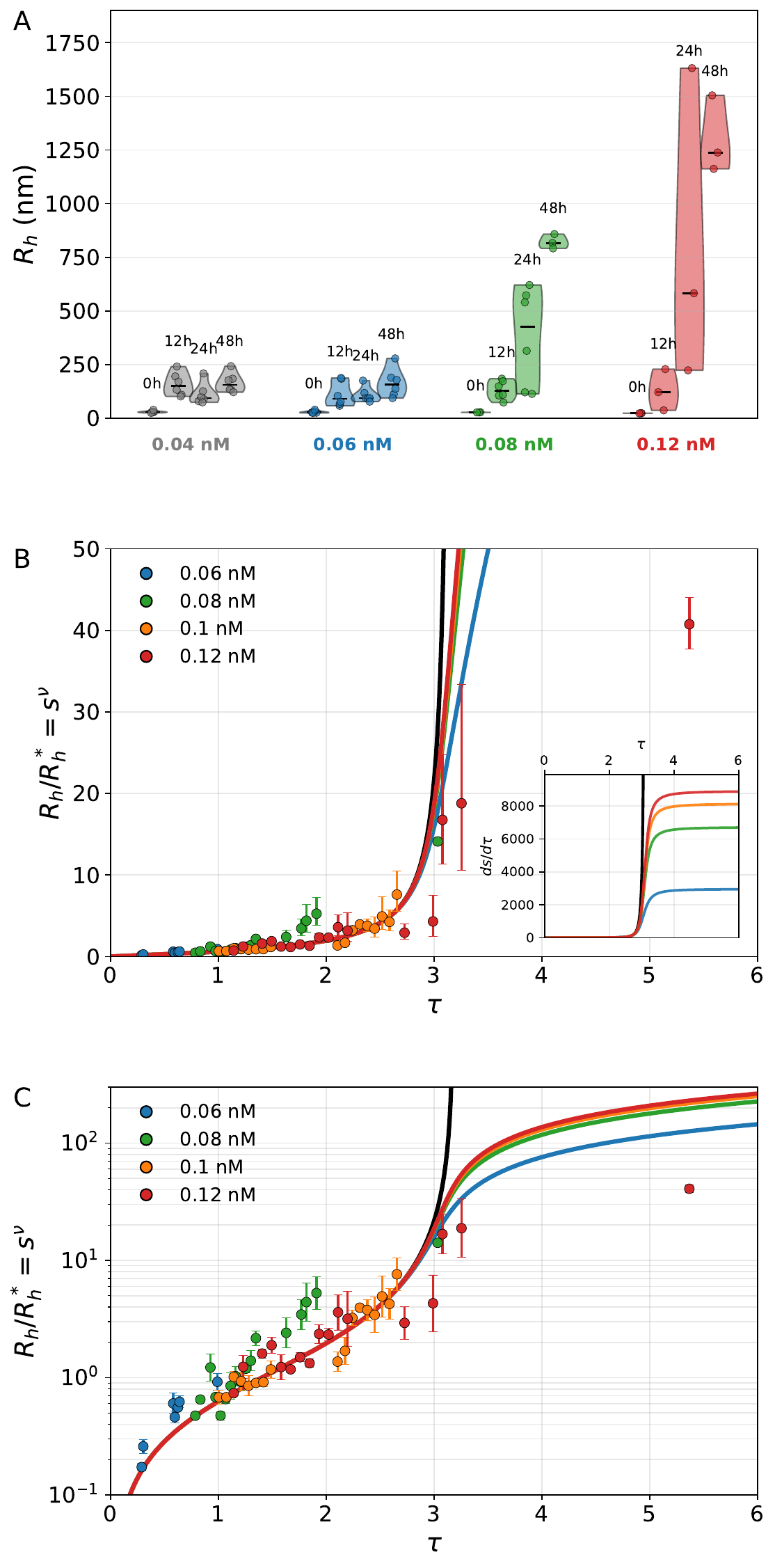}
\caption{DNA-linker aggregation of gold nanoparticles. 
(a) DLS hydrodynamic-radius distributions at fixed $C_p=0.04$ nM per complementary particle type for $C_l=0.04$--$0.12$ nM, measured at $t=0$, 12, 24, and 48 h. Low-$C_l$ samples arrest or remain near critical, whereas high-$C_l$ samples grow. 
(b) Rescaled kinetics, $R_h/R_h^*=s^\nu$ versus $\tau=t/t^*$, using $\nu=0.55$, $C_l^*=0.057$ nM, and $k_p^{-1}=16$ h; the solid line is the $\epsilon\to\infty$ linker-limited master curve. 
(c) Same data on a logarithmic scale, highlighting early acceleration and late-time deviations.}
\label{fig:exp}
\end{figure}

The supercritical branch contains two asymptotic regimes. In the formal linker-limited limit $\epsilon\to\infty$,
\begin{equation}
s(\tau)\simeq \left[(1-\nu)(\tau_0-\tau)\right]^{-1/(1-\nu)},
\label{eq:singularity}
\end{equation}
Here $\tau_0=\frac{\pi}{\sin(\pi\nu)} $. The apparent finite-time singularity expresses the positive feedback between mass growth and cluster reactivity. For any finite $\epsilon$, this acceleration is cut off when diffusion to a highly decorated cluster becomes rate limiting. At late times,
\begin{equation}
s\simeq \epsilon(\tau-\tau_0),
\qquad
R_h/R_h^*=s^\nu,
\label{eq:late}
\end{equation}
recovering Smoluchowski coarsening in reduced variables. Thus the same equation predicts early acceleration, crossover, and eventual sublinear growth of the experimentally relevant  hydrodynamic radius $R_h$.

We tested these predictions experimentally using  DNA-functionalized gold nanoparticles at fixed concentration $C_p=0.04$ nM  and variable linker concentration $C_l$ (Fig.~\ref{fig:exp}).  Hydrodynamic radii were measured by dynamic light scattering after the rapid initial linker-adsorption step. The data separate into two regimes under the same preparation protocol: $C_l=0.04$ and $0.06$ nM show arrested or near-critical growth over the experimental window, whereas $C_l=0.08$ and $0.12$ nM show sustained broadening and shifts to larger $R_h$.

A representative collapse of the growing and near-critical data is obtained with
\begin{equation}
C_l^*\simeq0.057~{\rm nM},
\qquad
k_p^{-1}\simeq16~{\rm h},
\label{eq:fit}
\end{equation}
using $\nu=0.55$. This value places the $C_l=0.06$ nM sample slightly above threshold, where $t^*$ is large and growth appears nearly arrested on a 48 h window. It also gives
\begin{equation}
\alpha\simeq1-\frac{C_p}{C_l^*}\simeq0.3,
\label{eq:alpha_fit}
\end{equation}
smaller than the naive single-contact estimate $\alpha\simeq1/2$, as expected for an effective excluded-surface parameter in aggregates with overlapping blocked regions and flexible linkers. The precise value of $C_l^*$ is not uniquely fixed by the available data, but the qualitative separation between linker-depleted arrest and self-sustained growth, together with the approximate scaling collapse, is robust. Late-time deviations for the largest aggregates are likely associated with sedimentation, spatial inhomogeneity, and other effects beyond the present well-mixed mean-field model.

In conclusion, we have shown theoretically and experimentally that linker-mediated aggregation exhibits kinetic criticality controlled by the linker-to-particle concentration ratio. Although related to percolation and gelation, this critical point is not an equilibrium transition but a bifurcation in kinetic fate. Below threshold, active linkers are depleted and growth arrests at finite cluster size; above threshold, linker conservation sustains continued coarsening. In the supercritical regime, redistribution of active linkers among fewer clusters transiently amplifies reactivity, producing self-accelerating growth until diffusion becomes rate limiting and restores classical Smoluchowski coarsening. Dynamic-light-scattering measurements on DNA-linked gold nanoparticles confirm the predicted separation between arrested and growing regimes and show approximate collapse under the theoretical scaling. This experimentally verified mechanism should be broadly relevant to DNA-programmed colloids, multivalent biomolecular binding, and aggregation-based sensing, where linker abundance, patch size, and activation barriers provide direct control over threshold amplification and late-time growth.

{\bf Acknowledgments}This research was conducted in part at the Center for Functional Nanomaterials, a U.S. Department of Energy Office of Science User Facility at Brookhaven National Laboratory, under Contract No. DE-SC0012704. The work was supported by the U.S. Department of Defense, Army Research Office, W911NF-22-2-0111. Z.A.A. was supported in part by the Human Frontier Science Program and the Zuckerman Israeli Postdoctoral Scholars Program.

\bibliography{main}

\end{document}

\title{Supplemental Material for ``Kinetic Criticality in Linker-Mediated Colloidal Aggregation''}
\author{Alexei V. Tkachenko}
\affiliation{Center for Functional Nanomaterials, Brookhaven National Laboratory, Upton, New York 11973, USA}
\author{Soojung Lee}
\affiliation{Department of Chemical Engineering, Columbia University, New York, New York 10027, USA}
\author{Zohar A. Arnon}
\affiliation{Department of Chemical Engineering, Columbia University, New York, New York 10027, USA}
\affiliation{Avram and Stella Goldstein-Goren Department of Biotechnology, Ben-Gurion University of the Negev, Beer Sheva 8410501, Israel}
\author{Oleg Gang}
\affiliation{Center for Functional Nanomaterials, Brookhaven National Laboratory, Upton, New York 11973, USA}
\affiliation{Department of Chemical Engineering, Columbia University, New York, New York 10027, USA}
\affiliation{Department of Applied Physics and Applied Mathematics, Columbia University, New York, New York 10027, USA}
\affiliation{Center for Nanomedicine, Institute for Basic Science and Department of Nano Biomedical Engineering, Yonsei University, Seoul 03722, Republic of Korea}
\date{\today}
\maketitle

This Supplemental Material gives the derivations, experimental details, and fitting conventions that are compressed in the Letter. It integrates the longer derivation from the earlier manuscript with the shorter PRL Supplemental Material entry, while using the notation of the revised Letter.

\section{Diffusion-limited reference kinetics}

Classical colloidal aggregation is described by the Smoluchowski picture, in which every encounter between diffusing clusters is reactive and growth is controlled solely by transport. A simplified kinetic equation for the cluster molar concentration $c(t)$ is
\begin{equation}
\dot{c}\approx -4\pi \cdot 10^{3} N_A \left\langle (D_i+D_j)(R_i+R_j)\right\rangle c^2\approx -\kappa_0 c^2 .
\label{eq:S_smol}
\end{equation}
Here $N_A$ is Avogadro's number, $D_i$ and $D_j$ are cluster diffusion coefficients, and $R_i$ and $R_j$ are hydrodynamic radii. For comparable clusters, the Einstein relation $D=k_BT/(6\pi\eta R)$ removes the explicit radius dependence and gives
\begin{equation}
\kappa_0=16\pi\cdot10^3N_A D r_p=\frac{8\cdot10^3RT}{3\eta} .
\label{eq:S_kappa0}
\end{equation}
Writing the mean cluster mass as
\begin{equation}
M(t)=\frac{C_p}{c(t)},
\label{eq:S_Mdef}
\end{equation}
with $C_p=c(0)$, Eq.~(\ref{eq:S_smol}) gives
\begin{equation}
M(t)=1+\kappa_0 C_p t .
\label{eq:S_Msmol}
\end{equation}
If the hydrodynamic radius scales with mass as
\begin{equation}
R_h\simeq r_p M^\nu,
\label{eq:S_RhM}
\end{equation}
then the classical diffusion-limited hydrodynamic growth law is $R_h\sim t^\nu$.

\section{Fast linker adsorption stage}

The Letter assumes that linker adsorption onto particles is fast compared with subsequent aggregation. Let $l\ll r_p$ denote the linker size. The diffusion-limited capture constant of a small linker by a particle of radius $r_p$ is
\begin{equation}
\kappa_l \simeq 4\pi\cdot 10^3 N_A D_l r_p,
\label{eq:S_kl}
\end{equation}
where $D_l$ is the linker diffusivity. Using Stokes--Einstein scaling,
\begin{equation}
\frac{D_l}{D}\simeq \frac{r_p}{l},
\label{eq:S_Dratio}
\end{equation}
and comparing Eq.~(\ref{eq:S_kl}) with Eq.~(\ref{eq:S_kappa0}), one obtains
\begin{equation}
\kappa_l\simeq \frac{r_p}{4l}\,\kappa_0 .
\label{eq:S_kl2}
\end{equation}
Therefore, the free-linker concentration decays on the time scale
\begin{equation}
\tau_l\simeq \frac{1}{\kappa_l C_p}=\frac{4}{\kappa_0 C_p}\frac{l}{r_p} .
\label{eq:S_taul}
\end{equation}
For short DNA linkers and nanoparticles, $l/r_p\ll1$, so this adsorption step is expected to precede the slower cluster aggregation stage.

\section{Active-linker conservation, threshold, and concentration conventions}

After the initial adsorption stage, aggregation proceeds by linker-mediated binding between clusters. Let $c_l(t)$ denote the concentration of active adsorbed linkers, i.e., linkers that remain available to form intercluster bridges. The number of mergers that have occurred by time $t$ is
\begin{equation}
N_{\rm merge}(t)=C_p-c(t).
\label{eq:S_merge}
\end{equation}
Every merger consumes one active linker. In addition, a merger creates a particle-particle contact that blocks part of the local surface from hosting additional active linkers. If $\alpha$ is the effective blocked surface fraction per contact, then the total number of linker adsorption sites lost in this way is
\begin{equation}
N_{\rm block}(t)=\alpha\frac{C_l}{C_p}\left[C_p-c(t)\right].
\label{eq:S_block}
\end{equation}
Thus
\begin{equation}
c_l(t)=C_l-\left[C_p-c(t)\right]-\alpha\frac{C_l}{C_p}\left[C_p-c(t)\right] .
\label{eq:S_clraw}
\end{equation}
Rearranging gives the conservation law
\begin{equation}
c_l(t)-\gamma c(t)=(1-\alpha)C_l-C_p,
\qquad
\gamma\equiv1+\alpha\frac{C_l}{C_p} .
\label{eq:S_balance}
\end{equation}
The kinetic threshold is obtained by setting the right-hand side of Eq.~(\ref{eq:S_balance}) to zero:
\begin{equation}
C_l^*=\frac{C_p}{1-\alpha} .
\label{eq:S_Clstar}
\end{equation}
The dimensionless distance from threshold is
\begin{equation}
\Lambda\equiv\frac{C_l/C_l^*-1}{\gamma},
\qquad
\gamma=1+\alpha\frac{C_l}{C_p} .
\label{eq:S_Lambda}
\end{equation}
With the characteristic mass scale
\begin{equation}
M^*=\frac{1}{|\Lambda|},
\label{eq:S_Mstar}
\end{equation}
it is natural to define
\begin{equation}
s\equiv |\Lambda|M,
\qquad M=\frac{s}{|\Lambda|} .
\label{eq:S_sdef}
\end{equation}
Using $M=C_p/c$ and Eq.~(\ref{eq:S_balance}), one obtains
\begin{equation}
c_l(s)=\gamma C_p |\Lambda|\frac{1\pm s}{s},
\label{eq:S_cls}
\end{equation}
where the upper sign corresponds to $C_l>C_l^*$ and the lower sign to $C_l<C_l^*$. In the subcritical branch, $c_l\to0$ as $s\to1$, giving arrested growth at $M=M^*$.

In the binary nanoparticle experiments, the reported particle concentration $C_p=0.04$ nM denotes the concentration of each complementary particle population. The same convention is used in the threshold estimate $\alpha=1-C_p/C_l^*$. The fact that only one of the two particle types is complementary to a given linker end is accounted for separately by the compositional factor $z=2$ in the effective reaction length derived below.

\section{Accessible surface concentration}

Consider a cluster containing $m$ particles. Each interparticle bond blocks a fraction $\alpha$ of a single-particle surface, so the accessible surface area of the cluster, measured in units of a single-particle surface area, is
\begin{equation}
A(m)=m-\alpha(m-1)=(1-\alpha)m+\alpha .
\label{eq:S_Am}
\end{equation}
If $c_m(t)$ is the concentration of clusters of mass $m$, the total accessible-surface concentration is
\begin{equation}
a(t)=\sum_m c_m A(m)=(1-\alpha)\sum_m m c_m+\alpha\sum_m c_m=(1-\alpha)C_p+\alpha c(t),
\label{eq:S_at}
\end{equation}
where $\sum_m m c_m=C_p$ and $\sum_m c_m=c(t)$. Dividing out the prefactor $(1-\alpha)$ defines the rescaled accessible-surface concentration
\begin{equation}
c_p(t)\equiv \frac{a(t)}{1-\alpha}=C_p+\frac{\alpha}{1-\alpha}c(t) .
\label{eq:S_cp}
\end{equation}
This quantity is not the literal particle concentration, which remains $C_p$, but a rescaled measure of the surface available for productive linker-mediated binding. In the scaling regime $M\gg1$, the correction term is small and $c_p(t)\simeq C_p$.

\section{Patch-to-particle reaction via electrostatic mapping}

To estimate the reaction-limited kernel, we consider the local diffusion problem for a complementary particle approaching a single active linker patch of effective radius $\xi_l$. The activation free energy for the patch reaction is included through
\begin{equation}
\xi_l\simeq l\exp(-w/k_BT),
\label{eq:S_xil}
\end{equation}
where $l$ is the linker length scale and $w$ is an effective barrier.

\begin{figure}[t]
\centering
\maybegraphics[width=\columnwidth]{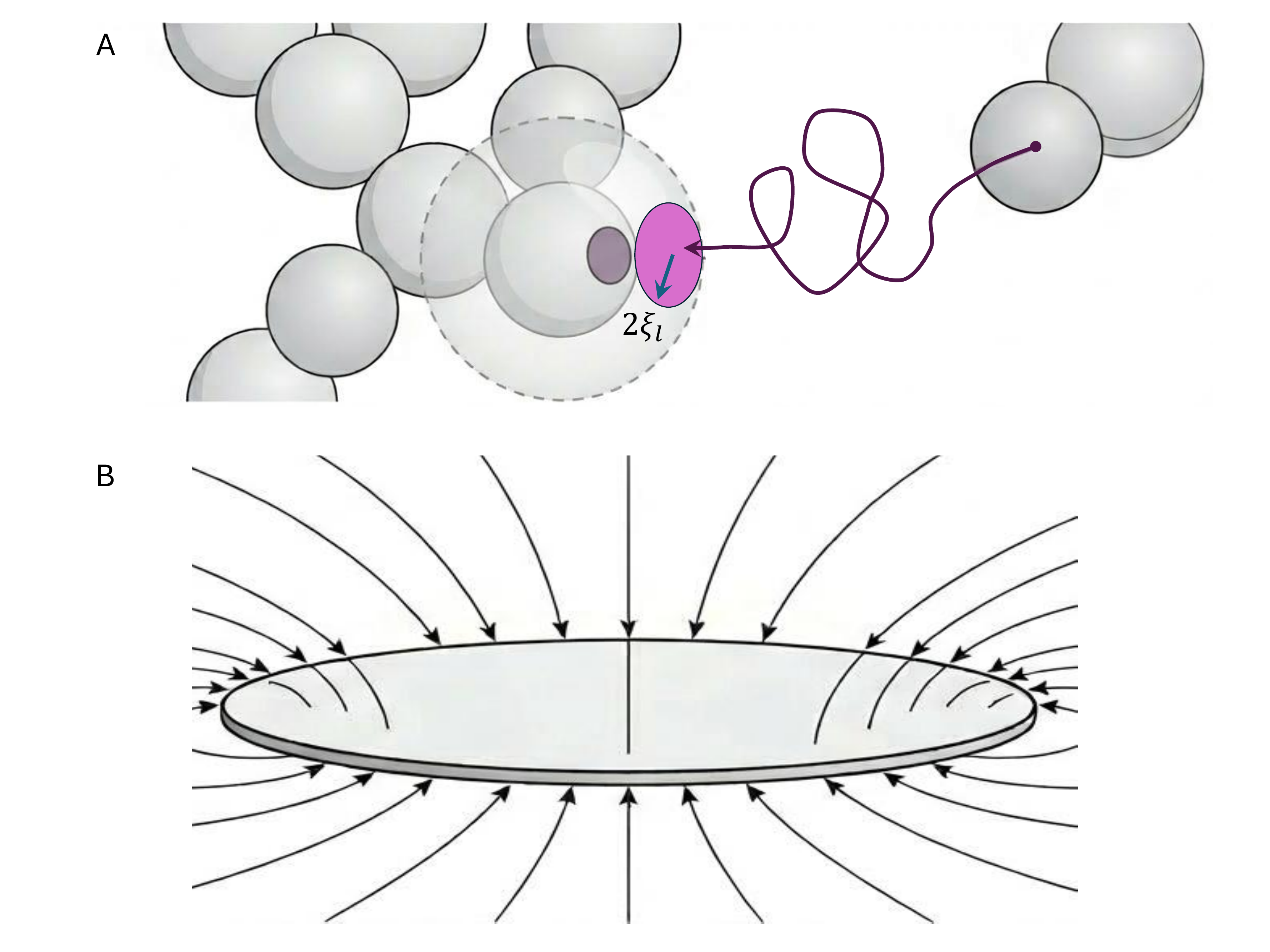}
\caption{Electrostatic mapping of patch-limited capture. (a) In the relative-coordinate representation, the approaching particle center diffuses near an excluded sphere of radius $2r_p$, and a physical linker patch of radius $\xi_l$ maps to an image patch of radius $2\xi_l$. For $\xi_l\ll r_p$, the surface is locally flat and the problem becomes diffusion to an absorbing disk in a reflecting plane. (b) After rescaling the normal coordinate to remove anisotropy, even reflection maps the half-space problem to the potential of a conducting disk in full space. The physical rate is the one-sided flux to the disk, one half of the mirrored two-sided flux.}
\label{fig:patch_mapping}
\end{figure}

In the relative-coordinate representation, the center of the incoming particle diffuses near an excluded sphere of radius $2r_p$, and the physical patch of radius $\xi_l$ maps to an image patch of radius $2\xi_l$ on that sphere. Since $\xi_l\ll r_p$, the excluded surface is locally flat on the scale of the patch. The local problem therefore reduces to diffusion in a half-space toward an absorbing disk embedded in an otherwise reflecting plane, as sketched in Fig.~\ref{fig:patch_mapping}.

Let $x$ denote the coordinate normal to the surface and $\bm{\rho}=(y,z)$ the tangential coordinates. The steady relative concentration field $u(x,\bm{\rho})$ obeys
\begin{equation}
D_\perp\partial_x^2u+D_\parallel\nabla_\rho^2u=0,
\qquad x>0,
\label{eq:S_aniso}
\end{equation}
with boundary conditions
\begin{align}
u(0,\rho)&=0, &&0\le \rho<2\xi_l,\label{eq:S_bc1}\\
\partial_xu(0,\rho)&=0, &&\rho>2\xi_l,\label{eq:S_bc2}\\
u(x,\rho)&\to u_\infty, &&x\to\infty .\label{eq:S_bc3}
\end{align}
The normal diffusivity is $D_\perp$ and the tangential diffusivity is $D_\parallel$. The anisotropy is removed by the rescaling
\begin{equation}
\tilde{x}=x\sqrt{\frac{D_\parallel}{D_\perp}},
\label{eq:S_xrescale}
\end{equation}
which transforms Eq.~(\ref{eq:S_aniso}) into the isotropic Laplace equation. The physical normal flux becomes
\begin{equation}
j_n=-D_\perp\partial_xu=-\sqrt{D_\perp D_\parallel}\,\partial_{\tilde{x}}u,
\label{eq:S_fluxrescale}
\end{equation}
so the original anisotropic problem is equivalent to isotropic diffusion with effective diffusivity
\begin{equation}
D_{\rm eff}=\sqrt{D_\perp D_\parallel} .
\label{eq:S_Deff}
\end{equation}

By even reflection across the plane $\tilde{x}=0$, the half-space problem maps to the electrostatic problem of a perfectly absorbing circular disk of radius $2\xi_l$ in full space. Using the diffusion-electrostatics correspondence,
\begin{equation}
k_{\rm disk}^{(2)}=4\pi D_{\rm eff}C_{\rm disk},
\label{eq:S_caprel}
\end{equation}
where the capacitance of a thin conducting disk of radius $a$ is $C_{\rm disk}=2a/\pi$. With $a=2\xi_l$,
\begin{equation}
k_{\rm disk}^{(2)}=16D_{\rm eff}\xi_l .
\label{eq:S_kpatch2}
\end{equation}
Only one side of the disk corresponds to physical flux from solution to the patch, so the relevant one-sided local capture rate is
\begin{equation}
k_{\rm patch}=8D_{\rm eff}\xi_l .
\label{eq:S_kpatch1}
\end{equation}

For a pair of clusters $i$ and $j$, this one-patch rate is multiplied by the number of productive patch-to-surface capture channels. Let $N_i$ be the number of active linkers on cluster $i$ and let ${\cal A}_i$ denote its accessible surface area measured in units of a single-particle surface. If only one out of $z$ particle types is complementary to a given linker end, then only the fraction $1/z$ of accessible surface is productive. The pairwise reaction-limited kernel scales as
\begin{equation}
\kappa_{ij}^{(r)}\sim\frac{8\xi_l}{z}(D_i+D_j)\left(N_i{\cal A}_j+N_j{\cal A}_i\right),
\label{eq:S_kij}
\end{equation}
where $z=1$ for a one-component system and $z=2$ for the binary complementary system used experimentally. In passing from $D_{\rm eff}$ to $D_i+D_j$, order-unity geometric factors are absorbed into the effective reaction length.

Averaging Eq.~(\ref{eq:S_kij}) over the cluster population and using
\begin{equation}
\langle N\rangle=\frac{c_l}{c},
\qquad
\langle {\cal A}\rangle=(1-\alpha)\frac{c_p}{c},
\label{eq:S_NAavg}
\end{equation}
one obtains
\begin{equation}
\kappa_r\simeq\kappa_0\frac{\xi}{R_h}\frac{c_l(t)c_p(t)}{c^2} .
\label{eq:S_kr0}
\end{equation}
The blocked-surface factor, the compositional factor, and the local electrostatic prefactor are absorbed into
\begin{equation}
\xi=\frac{2(1-\alpha)}{\pi z}\,\xi_l\simeq\frac{2(1-\alpha)}{\pi z}\,l\exp(-w/k_BT) .
\label{eq:S_xi}
\end{equation}
In the scaling regime $M\gg1$, $c_p(t)\simeq C_p$, giving the main-text form
\begin{equation}
\kappa_r\simeq\kappa_0\frac{\xi}{R_h}\frac{c_l(t)C_p}{c^2} .
\label{eq:S_kr_main}
\end{equation}

\section{Reduced kinetic equation}

The full aggregation rate is written as the harmonic sum of the diffusion-limited and reaction-limited resistances,
\begin{equation}
\dot c=-\frac{c^2}{\kappa_0^{-1}+\kappa_r^{-1}} .
\label{eq:S_full_rate}
\end{equation}
Using Eq.~(\ref{eq:S_kr_main}) with
\begin{equation}
R_h=r_pM^\nu=r_p|\Lambda|^{-\nu}s^\nu,
\qquad
c=\frac{C_p|\Lambda|}{s},
\label{eq:S_Rh_c}
\end{equation}
and substituting Eq.~(\ref{eq:S_cls}), one finds
\begin{equation}
\frac{c_l C_p}{c^2}=\gamma\frac{s(1\pm s)}{|\Lambda|} .
\label{eq:S_clc}
\end{equation}
Therefore,
\begin{equation}
\kappa_r=\kappa_0\frac{\xi\gamma}{r_p}|\Lambda|^{\nu-1}s^{1-\nu}(1\pm s) .
\label{eq:S_krs}
\end{equation}
Since $s=|\Lambda|C_p/c$, Eq.~(\ref{eq:S_full_rate}) becomes
\begin{equation}
\dot s=\kappa_0C_p|\Lambda|\left(1+\frac{\epsilon s^{\nu-1}}{1\pm s}\right)^{-1},
\label{eq:S_sdot}
\end{equation}
where
\begin{equation}
\epsilon=\frac{r_p}{\xi\gamma}|\Lambda|^{1-\nu} .
\label{eq:S_eps}
\end{equation}
Introduce
\begin{equation}
t^*\equiv\frac{1}{k_p\gamma|\Lambda|^\nu}=\frac{r_p}{\xi\kappa_0C_p\gamma|\Lambda|^\nu},
\qquad
k_p\equiv\frac{\xi\kappa_0C_p}{r_p},
\label{eq:S_tstar}
\end{equation}
and $\tau=t/t^*$. Then
\begin{equation}
\frac{ds}{d\tau}=\left(\frac{s^{\nu-1}}{1\pm s}+\frac{1}{\epsilon}\right)^{-1} .
\label{eq:S_scaled}
\end{equation}
The upper sign gives unbounded supercritical growth. The lower sign gives a fixed point at $s=1$, corresponding to arrested growth at $M=M^*$.

\section{Exact implicit solutions and asymptotic limits}

Equation~(\ref{eq:S_scaled}) may be written as
\begin{equation}
\frac{d\tau}{ds}=\frac{1}{\epsilon}+\frac{s^{\nu-1}}{1\pm s} .
\label{eq:S_dtauds}
\end{equation}
For the supercritical branch, integration gives
\begin{equation}
\tau_+(s)=\frac{s}{\epsilon}+\int_0^s\frac{u^{\nu-1}}{1+u}\,du
=\frac{s}{\epsilon}+B_{s/(1+s)}(\nu,1-\nu),
\label{eq:S_tauplus_int}
\end{equation}
where
\begin{equation}
B_z(a,b)\equiv\int_0^z t^{a-1}(1-t)^{b-1}\,dt
\label{eq:S_Bdef}
\end{equation}
is the incomplete beta function. For the subcritical branch,
\begin{equation}
\tau_-(s)=\frac{s}{\epsilon}+\int_0^s\frac{u^{\nu-1}}{1-u}\,du,
\qquad 0\le s<1 .
\label{eq:S_tauminus_int}
\end{equation}
Equivalently,
\begin{equation}
\tau_-(s)=\frac{s}{\epsilon}+\frac{s^\nu}{\nu}\,{}_2F_1(\nu,1;\nu+1;s),
\qquad 0\le s<1,
\label{eq:S_tauminus_hyper}
\end{equation}
which is the incomplete-beta representation $B_s(\nu,0)$ understood through analytic continuation or through the explicit integral in Eq.~(\ref{eq:S_tauminus_int}). Unlike the supercritical branch, $\tau_-(s)$ diverges logarithmically as $s\to1^-$.

For a system prepared at a finite initial mass $M(0)$, the measured time is shifted by the initial reduced mass
\begin{equation}
s_0=|\Lambda|M(0) .
\label{eq:S_s0}
\end{equation}
Thus
\begin{equation}
\frac{t_{\rm phys}}{t^*}=\tau_\pm(s)-\tau_\pm(s_0) .
\label{eq:S_t_vs_tau}
\end{equation}
For a monomeric initial state, $M(0)=1$ and $s_0=|\Lambda|$.

For large $s$, Eq.~(\ref{eq:S_tauplus_int}) gives
\begin{equation}
\tau_+(s)=\frac{s}{\epsilon}+B(\nu,1-\nu)-\frac{s^{\nu-1}}{1-\nu}+O\left(s^{\nu-2}\right),
\label{eq:S_tauplus_asym}
\end{equation}
where the complete beta function is
\begin{equation}
\tau_0\equiv B(\nu,1-\nu)=\frac{\Gamma(\nu)\Gamma(1-\nu)}{\Gamma(1)}=\frac{\pi}{\sin(\pi\nu)} .
\label{eq:S_tau0}
\end{equation}
In the formal linker-limited limit $\epsilon\to\infty$, the linear term in Eq.~(\ref{eq:S_tauplus_asym}) drops out, giving the apparent finite-time singularity
\begin{equation}
\tau_0-\tau\simeq\frac{s^{\nu-1}}{1-\nu},
\qquad
s(\tau)\simeq\left[(1-\nu)(\tau_0-\tau)\right]^{-1/(1-\nu)} .
\label{eq:S_super_sing}
\end{equation}
For finite $\epsilon$, the term $s/\epsilon$ eventually dominates and cuts off the singularity. At late times,
\begin{equation}
s(\tau)\simeq\epsilon(\tau-\tau_0) .
\label{eq:S_super_linear}
\end{equation}
For the subcritical branch, as $s\to1^-$,
\begin{equation}
\tau_-(s)=-\ln(1-s)+{\rm const}+O(1-s),
\label{eq:S_sub_log}
\end{equation}
so
\begin{equation}
s(\tau)\simeq1-Ae^{-\tau},
\label{eq:S_sub_exp}
\end{equation}
where $A$ depends on the initial condition.

\section{Preparation of DNA-functionalized gold nanoparticles and DLS measurements}

Thiolated oligonucleotides purchased from Integrated DNA Technologies were first reduced by tris[2-carboxyethyl] phosphine (TCEP) in distilled water at a molar ratio of 1:100 (oligonucleotide:TCEP). The reduced oligonucleotides were conjugated to 20 nm gold nanoparticles (Ted Pella, \#15705-1) through Au--S bonds at a DNA:AuNP molar ratio of 1200:1 and incubated for 1.5 h. Sodium dodecyl sulfate was then added to a final concentration of 0.01\% to prevent aggregation, and the solution was buffered with 10 mM phosphate buffer at pH 7 for 1.5 h.

Sodium chloride was introduced in five incremental additions to a final concentration of 0.3 M NaCl. After each addition, the solution was sonicated for 10--15 s and incubated for 50 min, followed by overnight incubation at room temperature. Excess reagents were removed by centrifugation and sequential washing with 10 mM phosphate buffer and 0.1 M NaCl, followed by three additional washes with water.

Particle concentrations were determined by UV--Vis absorption, and each complementary AuNP population was diluted to a final concentration of 40 pM. The single-stranded DNA linker sequence was prepared as a 1 nM stock solution and added to the AuNP mixture to achieve the targeted linker concentrations $C_l=0.04$--$0.12$ nM. At each measurement time, samples were briefly vortexed and the hydrodynamic cluster size was measured by dynamic light scattering using a Zetasizer Nano-ZS instrument (Malvern Instruments, U.K.).

\section{Fitting parameters and corrected notation}

The experimental scaling used
\begin{align}
R_h&\simeq r_pM^\nu=R_h^*s^\nu,\label{eq:S_fit_Rh}\\
t^*&=\frac{1}{\gamma k_p|\Lambda|^\nu},\label{eq:S_fit_t}\\
\epsilon&=\frac{r_p}{\xi\gamma}|\Lambda|^{1-\nu}=\kappa_0C_pt^*|\Lambda| .\label{eq:S_fit_eps}
\end{align}
The exponent was fixed at $\nu=0.55$. A representative global fit gives
\begin{equation}
C_l^*=0.057~{\rm nM},
\qquad
k_p^{-1}=16~{\rm h} .
\label{eq:S_fit_values}
\end{equation}
This implies
\begin{equation}
\alpha=1-\frac{C_p}{C_l^*}\simeq0.30,
\label{eq:S_alpha_fit}
\end{equation}
using the per-type particle concentration $C_p=0.04$ nM.

Table S1 lists the parameters generated by this fit. To avoid ambiguity, the table distinguishes the unnormalized reduced linker excess,
\begin{equation}
\Delta\equiv\frac{C_l}{C_l^*}-1,
\label{eq:S_delta}
\end{equation}
from the scaled control parameter used in the theory,
\begin{equation}
\Lambda=\frac{\Delta}{\gamma} .
\label{eq:S_lambda_delta}
\end{equation}
This distinction is important because $M^*=1/|\Lambda|$ and $t^*=1/(\gamma k_p|\Lambda|^\nu)$ use $\Lambda$, not $\Delta$.

\begin{center}
\begin{minipage}{0.98\columnwidth}
\textbf{Table S1.} Representative fitting parameters for $C_l^*=0.057$ nM, $C_p=0.04$ nM per particle type, $\alpha=0.30$, $\nu=0.55$, and $k_p^{-1}=16$ h. The column $\Delta$ is the unnormalized excess linker concentration, while $\Lambda=\Delta/\gamma$ is the control parameter used in the scaling theory.
\end{minipage}
\begin{ruledtabular}
\begin{tabular}{ccccccc}
$C_l$ (nM) & $\Delta$ & $\gamma$ & $\Lambda$ & $t^*$ (h) & $M^*$ & $r_p$ (nm) \\
\hline
0.04 & $-0.30$ & 1.30 & $-0.230$ & 27.68 & 4.35 & 29.24 \\
0.06 & $+0.053$ & 1.45 & $+0.036$ & 68.42 & 27.50 & 27.14 \\
0.08 & $+0.40$ & 1.60 & $+0.253$ & 21.35 & 3.96 & 27.32 \\
0.10 & $+0.75$ & 1.75 & $+0.432$ & 14.54 & 2.31 & 33.50 \\
0.12 & $+1.11$ & 1.89 & $+0.583$ & 11.36 & 1.71 & 23.60 \\
\end{tabular}
\end{ruledtabular}
\end{center}

Comparable collapses can be obtained over an approximate interval $C_l^*=0.04$--$0.07$ nM by refitting $k_p$. The quantitative value of $C_l^*$ should therefore not be overinterpreted. The robust experimental conclusion is the separation between low-linker arrested or near-critical behavior and high-linker self-sustained growth, together with approximate scaling of the growing datasets.

\section{Model limitations}

The theory is a homogeneous mean-field description of irreversible, well-mixed aggregation. It does not explicitly include reversible DNA unbinding, bond exchange, aggregate restructuring, many-body hydrodynamics, sedimentation-driven spatial inhomogeneity, or the full cluster-size distribution. These effects are expected to be most important at late times and for the largest aggregates. In particular, gravitational sedimentation can deplete large clusters from the bulk probed by dynamic light scattering, making the observed late-time growth slower than the homogeneous prediction. Such effects do not alter the central kinetic mechanism: active-linker conservation produces a threshold between linker-depleted arrest and self-sustained aggregation, while patch-limited reactivity produces the early accelerating regime and diffusion-limited transport cuts it off.